\documentclass{iaus}
\usepackage{graphicx}
\def\HI{H\,{\sc i}}
\newcommand{\kms}{$\,$km$\,$s$^{-1}$}

\newcommand{\msun}{{$M_\odot$}}

\title[Gas Outflows in Young Radio AGN] 
{Cold and Warm Gas Outflows in Radio AGN}

\author[R. Morganti et al.]   
{Raffaella Morganti$^{1,2}$, Joanna Holt$^{3}$, \\ 
Clive Tadhunter$^{4}$, \and Tom Oosterloo$^{1,2}$
}

\affiliation{$^1$ Netherlands Institute for Radio Astronomy,\\ Postbus 2,
7990 AA, Dwingeloo, The Netherlands\\
Email: {\tt morganti@astron.nl} \\[\affilskip]
$^2$ Kapteyn Astronomical Institute, University of Groningen, \\ P.O. Box 800,
9700 AV Groningen, The Netherlands \\ [\affilskip]
$^3$  Leiden Observatory, Leiden University, PO Box 9513, 2300 RA
  Leiden, The Netherlands  \\ [\affilskip]
$^4$ Department of Physics and Astronomy, University of Sheffield,
  Sheffield, S3 7RH, UK}

\pubyear{2010}
\volume{267}  
\pagerange{1--9}
\jname{Co-Evolution of Central Black Holes and Galaxies}
\editors{B.M.\ Peterson, R.S.\ Somerville, \& T.\ Storchi-Bergmann, eds.}
\begin{document}

\maketitle

\begin{abstract}
The study of the conditions and the kinematics of the gas in the
central region of AGN provides important information on the relevance
of feedback effects connected to the nuclear activity. Quantifying these
effects is key for constraining galaxy evolution models.  
Here we  present a  short summary of our recent efforts  to study the
occurrence and the impact of gas outflows in radio-loud AGN that are
in their first phase of their evolution.  Clear evidence for AGN-induced
outflows have been found for the majority of these young radio
sources. The outflows are detected both in (warm) ionized as well in
(cold) atomic neutral gas and they are  likely to be  driven (at least
in most of the cases) by the interaction between the expanding jet and
the medium.  
The mass outflow rates of the cold gas (\HI) appear to be systematically
higher than those of the ionized gas. The former reach up to $\sim 50$
\msun\ yr$^{-1}$, and are in the same range as  ``mild''
starburst-driven superwinds in ULIRGs, whilst the latter are currently
estimated to be a few solar masses per year. 
However, the kinetic powers associated with  these gaseous outflow are a relatively small fraction (a few 
$\times 10^{-4}$) of the Eddington luminosity of the galaxy. Thus,
they do not appear to match the requirements of the galaxy evolution
feedback models.


\end{abstract}

\firstsection 
\section{Introduction}

Relativistic plasma jets represent one of the possible ways in which
the enormous energy released by an active nucleus can interact with and
affect the interstellar medium. In powerful radio sources, the kinetic
power of such jets is comparable to the energy released via
radiation emission or  winds and, therefore, it is important to
consider the impact that this can have on the life and evolution of
the host galaxy.  
Quantifying this impact is necessary in order to set tighter constraints 
on the effects of feedback from AGN. Indeed, due to a lack of such
observational constraints, feedback is often invoked  as a ``black box''
and often relies on powerful AGN winds  (see, e.g., Di Matteo  et
al. 2005; Hopkins et al. 2005). Whilst this may be true for
radio-quiet AGN, it is likely that in radio-loud AGN the expanding
radio jets also contribute to the feedback via  jet-induced outflows.  

Different diagnostics have already clearly shown that, on scales of
tens to hundreds of kiloparsecs, radio emission can have an impact on the
ISM (e.g., on its kinematics and possibly ionization and
morphology). For example, in many powerful radio galaxies, the
extended emission lines show highly perturbed kinematics in the
regions along the radio axis  (see some of the nearby examples
shown by, e.g., Clark
et al. 1997, Villar-Mart\'{\i}n et al. 1999, and Sol\'orzano-I\~narrea et
al. 2001).  
In high-$z$ radio galaxies, high velocity gradients are observed in
the Ly$\alpha$ halos  (see, e.g., Villar-Mart\'{\i}n et al. 2001 and
references therein). In some cases, the presence of fast gas outflows ($ >
1000$\,\kms) has been confirmed through integral field spectroscopy
(see Nesvadba et al. 2006, 2009). Extreme examples of interaction
between the radio plasma and ISM are the X-ray cavities in the hot gas
that are hollowed out by expanding radio lobes (see, e.g., McNamara
\& Nulsen 2007). The existence and the characteristics of such cavities shows that  the radio
emission can prevent the hot gas from cooling and forming stars.  
These studies already illustrate that the impact of radio plasma jets
is not only limited to their (often) very collimated structures, but
that the effect extends to the surrounding  gas by the
impinging of the jet in the ISM.  
This perturbed, wide region is also seen in the results of numerical
simulations (see, e.g.,  Krause et al. 2009 and references therein for some recent examples)
showing how the passage of the jet creates a large
cocoon of shocked ambient gas.  
  
\begin{figure}[b]
\begin{center}
 \includegraphics[width=5.3in]{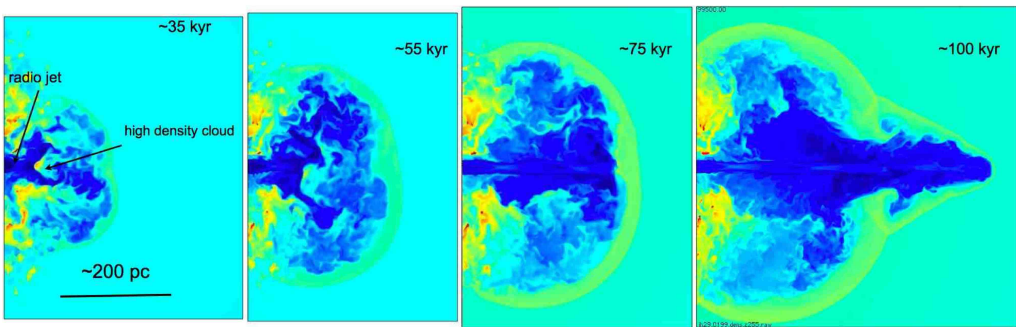} 
 \caption{Snapshots of numerical simulations illustrating, at
 different times, the evolution of the interaction of a plasma jet
 with an inhomogeneous, clumpy medium (Sutherland \& Bicknell 2007; Wagner
 \& Bicknell in preparation).}
   \label{fig1}
\end{center}
\end{figure}

This interaction  is likely to be even more relevant during the early
stages of the evolution of the radio source, i.e., when the jet is
interacting with the natal cocoon of gas deposited in the inner
regions during the triggering event, often considered  a  merger
and/or interaction (see, e.g., Bicknell, Dopita, \& O'Dea 1997).
Figure\,\ref{fig1}  illustrates  
examples of the evolution with time of the highly complex interplay
between a newly born radio jet and the rich and clumpy surrounding
ISM. These are obtained from  the numerical simulations of Geoff
Bicknell and collaborators  (see Sutherland \& Bicknell 2007 and Saxton et
al.  2005 and references therein).  

This short summary presents our recent efforts to quantify, using
observations of ionized and neutral atomic hydrogen,  the
characteristics of the gas in the nuclear regions of young radio
sources and the occurrence and the   
impact of gas outflows (and in particular  jet-driven outflows)
connected to the first phase of the evolution of their radio jets.

\begin{figure}[b]
\begin{center}
 \includegraphics[width=5in]{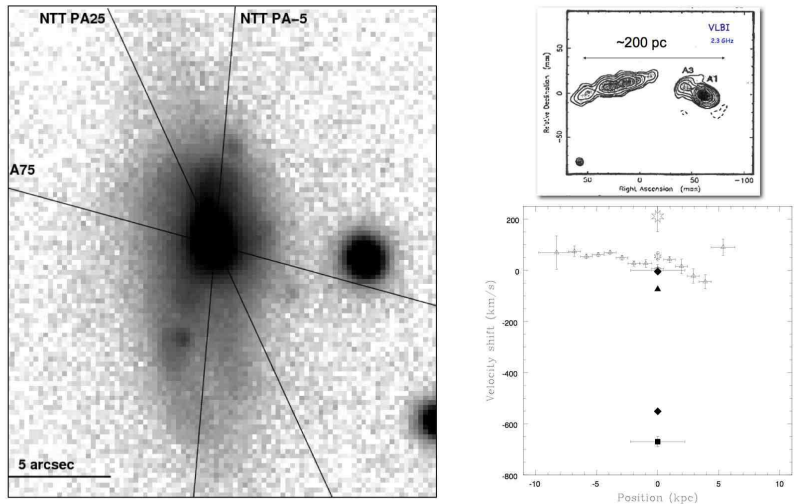} 
 \caption{{\em Left:} Deep VLT $r$-image of PKS1549--79 
($5'' = 12$\, kpc). The slit positions used for the long-slit spectroscopic
 observations are shown for reference.  
 {\em Right top:} VLBI  2.3\,GHz image of PKS~1549--79. The stronger component
 represents the radio core. 
{\em Right bottom:} Radial velocity profiles  of
 PKS 1549--79 obtained from the slit position PA-5. Small open  and
 filled triangles represent the narrow and intermediate components of
 H$\alpha$ respectively.  
Overplotted is the radial velocity of the deep \HI\ 21\,cm  absorption
(large filled triangle at $ -30$\,km\, s$^{-1}$). 
See Holt et al. (2006) for more details.  
}
  \label{fig2}
\end{center}
\end{figure}

\section{\boldmath Young Radio Sources: The Ionized Gas $\ldots$}

Newly born or young radio sources are  identified by their small linear
size combined with steep radio spectra. They are usually dubbed 
``gigahertz-peaked'' or ``compact steep spectrum'' sources (see Fanti et
al. 1995 and O'Dea 1997 for reviews). We have used a sample of 14
powerful, young radio sources (Holt et al. 2008, 2009)  to investigate, from
the very beginning of their evolution, the impact of the radio plasma
on the surrounding gas. 
Young radio sources are known to often live  in a gas-rich medium as
expected if a merger or interaction has provided the triggering fuel for
the AGN (see, e.g., de Vries et al. 1999; Axon  et al. 2000; Vermeulen
et al. 2003; Morganti et al. 2005).   
Indeed, this is further supported by our recent results which show
that both the ionized gas and neutral atomic hydrogen
(\HI) are commonly observed in these sources.  

In the case of the ionized gas, the observed emission lines have been
modeled in order to investigate the kinematics and ionization
properties of the gas (see Holt, Tadhunter, \& Morganti 2008, 2009 for
details).  To enable accurate line modeling, the continuum
(comprised of a derived nebular continuum and a modeled stellar
continuum) was subtracted.  This modeling has also allowed us to
investigate the stellar populations in these sources, yielding
particularly interesting results (Holt et al. 2007). Most of these
young radio sources contain a significant young ($ < 0.5$\,Gyr to down
to a few Myr) stellar population that accounts for a significant
fraction of the stellar mass ($\sim$10--50\%). This further suggests
that these sources have gone through a gas-rich merger event in their
recent past and this event is likely responsible for both the burst of
star formation and the triggering of the AGN.

\begin{figure}[b]
\begin{center}
 \includegraphics[width=4in]{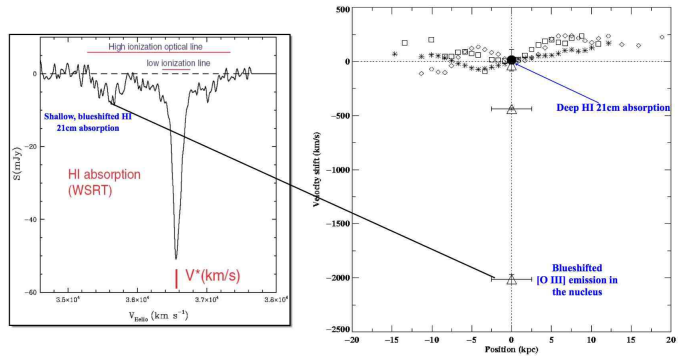} 
 \caption{{\em Left:} \HI\ profile  showing a deep absorption 
   close to the systemic velocity and a broad blueshifted wing
   ranging more than 1000 \kms. {\em Right:} Radial velocity profiles  of
   PKS~1345+12.  For more details, see  Holt et
   al. (2003). Overplotted is the radial velocity of the deep
   \HI\ 21cm  absorption and the location of the 
blueshifted absorption is also indicated.
}
  \label{fig3}
\end{center}
\end{figure}

We find that the ionized gas often extends  well beyond the radio
source (up to $\sim\! 20$\, kpc). These halos usually have 
quiescent kinematics (sometimes suggesting that the gas is in
rotation), in agreement with the fact that they have not been
affected by the passage of the radio source.  The presence of such
halos is shown in Figure\,\ref{fig2} and Figure\,\ref{fig3}  in the cases
of PKS~1549--79 and PKS~1345+12 respectively. Thus, we have used the
kinematics of the quiescent halos of ionized gas  to derive the systemic
velocity of the host galaxy.  
Deriving a reliable systemic velocity  has been a key step in the
interpretation of the kinematics of the gas (both warm and cold). 

On the other hand, the emission lines in the nuclear regions are
typically very complex (confirming earlier findings, e.g.,  Gelderman
\& Whittle 1994), requiring multiple Gaussian components to model
them.  Furthermore, the majority of the objects studied show that the
broader components tend to be blueshifted compared to the systemic
velocity, suggesting nuclear outflows. The shifts are up to
2000\,\kms\ (see, e.g., PKS~1345+12 in Figure\,\ref{fig3}, Holt et
al. 2003). These results show that the {\em presence of fast gas
outflows is (almost) ubiquitous in these objects}.  Although limited
by small number statistics, Holt et al. (2008) noted that the
characteristics of these fast gaseous outflow appear to be also
related to the size and to the orientation (with respect to the line
of sight) of the source. In particular, the most extreme of these
nuclear outflows are in the smaller gigahertz-peaked sources (with
linear size $< 1$\,kpc).

As final remark before moving to the characteristics of the cold gas,
it is interesting to note that  the hot component of the gas also shows
signatures of strong interactions between the radio jets/lobes and the
ISM. 
For example,  {\em Chandra} has revealed how even the lowest luminosity
radio sources will be initially overpressured, expanding
supersonically and shock heating their host galaxy ISM (e.g., Heinz et
al. 1998). These effects are observed in a number of objects (see,
e.g., Jetha et al. 2008, Croston et al. 2007, and Massaro et al. 2009), the
most famous being Centaurus A (Kraft et al. 2003; Croston et
al. 2009). 

\section{\boldmath $\ldots$ and the H\,{\small I}}

 \HI\  absorption against the strong radio continuum is often detected
 in the nuclear regions of  radio galaxies, but it is even more  common
 in powerful, young radio sources (see also Vermeulen et al. 2003).
 Although the detection of \HI\ absorption  gives only a limited view
 of the distribution of the cold atomic gas (being limited to the
 regions in front of the radio continuum), it still represents a
 sensitive tool to detect gas with column densities down to  
$N_{\mbox{{\scriptsize H}{\tiny I}}} \sim 10^{19}$\,cm$^{-2}$,  
much lower than what can be traced
 with \HI\ in emission, especially in distant  objects.  
 Different structures can produce the observed \HI\ absorption
 (e.g., circumnuclear disk, infalling clouds perhaps connected with the
 fueling of the AGN,  outflowing gas or large-scale disks seen in
 projection against the nuclear regions of the galaxy). In order to
 disentangle which of these structures is actually responsible for the
 absorption,  it is  crucial to combine the \HI\ results with those
 from the ionized gas, particularly for establishing a reliable
 systemic velocity.

\begin{figure}[b]
\begin{center}
\includegraphics[width=1.8in]{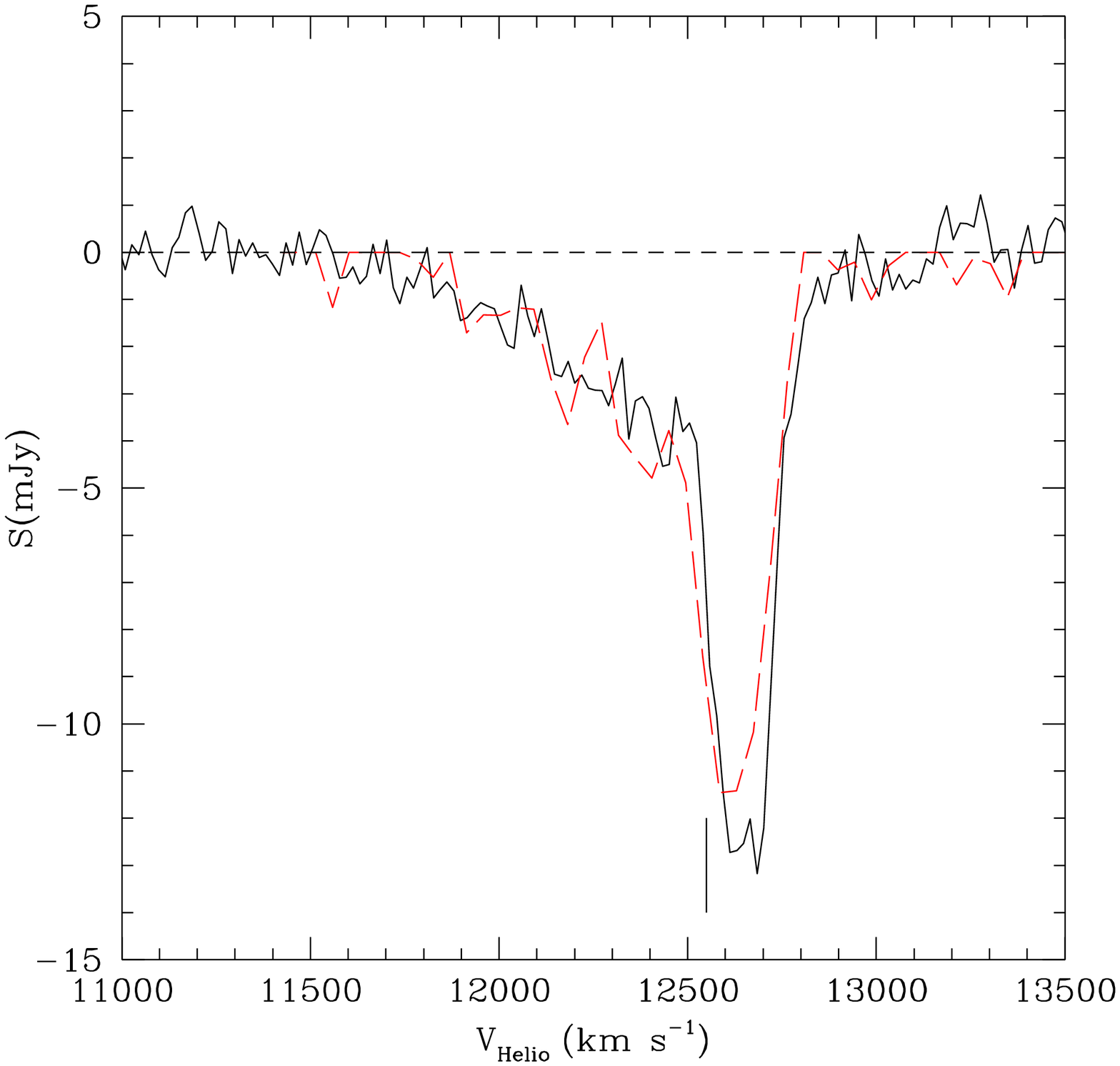} 
\includegraphics[width=1.8in]{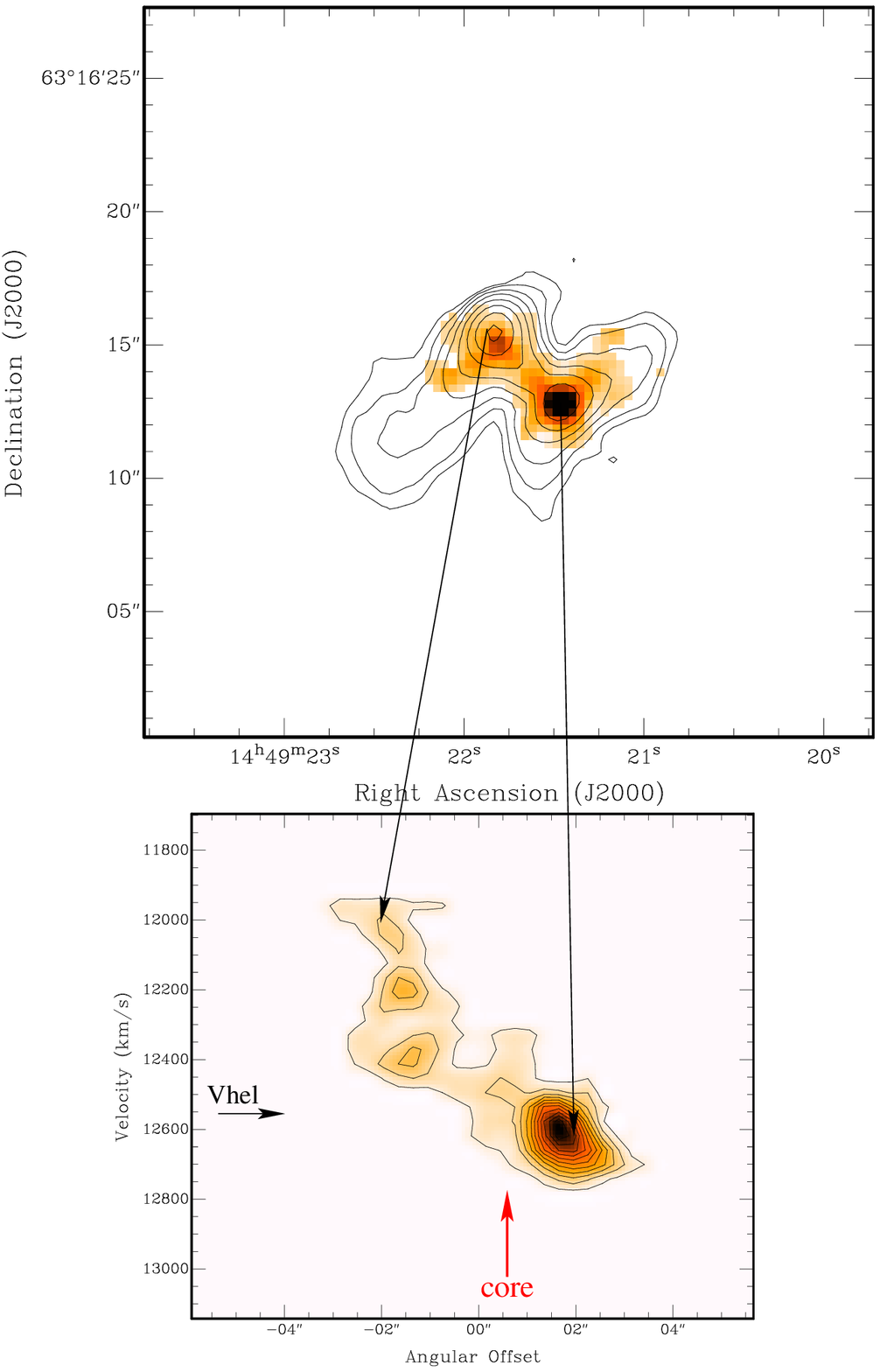}
  
\caption{{\em Left:} \HI\ absorption  profile obtained with the WSRT
  (solid line) superimposed on the integrated spectrum from the 
high-spatial resolution VLA observations (long-dashed). The profile
  shows a deep, relatively narrow, absorption and a broad component
  that covers more than 1000\,\kms\ at zero intensity,  mostly
  blueshifted compared to the systemic 
velocity indicated by the vertical solid line.  
{\em Right:} Panel showing 
the radio continuum image in contour 
and the integrated \HI\ absorption
(grey scale) from the VLA data {\em (top)} 
and the  position-velocity
plot from a slice passing 
through the two lobes {\em (bottom)}.  
The gray scale image represents the total
intensity of the \HI\ absorption, see Morganti et al. (2005b) for
details.} 
\  \label{fig4}
\end{center}
\end{figure}

A surprising result in the study of \HI\ in young radio sources
has been the discovery of fast \HI\ outflows (Morganti et al. 2003,
2005a and references therein). These outflows are characterized by the
detection of broad, blueshifted \HI\ absorption profiles (see, for
example, the case of PKS~1345+12 in Figure\,\ref{fig3} and of 3C~305 in
Figure\,\ref{fig4}).  
The velocities of these outflows exceed, in the most extreme cases,
1000\,\kms. Such high velocities were not expected to be  associated
with cold gas. The optical depth of the outflowing \HI\  gas component
is typically quite low, therefore making these absorption structures
particularly difficult to detect with present-day radio
telescopes.    

It is also difficult  to identify the location of the fast outflowing
gas. This is the main limitation if we want to understand their
origin.  Various possible outflow driving  mechanisms include
starburst superwinds, 
quasar-induced winds, and  expanding radio jets.  
Most of the outflows are limited to the central regions ($< $ a few 
kpc), i.e., co-spatial with the radio source.  This rules out the wide
starburst wind as the origin (see, e.g., Batcheldor et al. 2007 for more
details). 
The radio/optical alignment and the co-spatial location between the
outflows in ionized and neutral gas  and the radio structure (in
particular radio hot spots) is instead a  strong indication that
expanding radio jets could be a more likely mechanism. Indeed, it is
interesting to see that in the few cases where the location of the
\HI\ outflow has been identified, it is off-nucleus and coincident
with regions where the jet appears to strongly interact with the
surrounding medium.   
One of the best cases is the radio galaxy  3C~305 (Morganti et al. 2005b),
shown in Figure\,\ref{fig4}.
The broad \HI\ absorption is found $\sim 1.6$ kpc off-nucleus at the
location of the northeast radio lobe. This is the  same location where an
outflow of ionized gas occurs (Jackson et al. 1995). 
Further support for this hypothesis comes from the
{\em Chandra} detection of hot, X-ray emitting gas (Massaro et al. 2009)
that has been interpreted as  thermal emission due to the  interaction
between matter outflows and shock-heated environment gas. 
The fact that the two phases of the gas (cold neutral and warm
ionized) show similar kinematics, suggests that these  outflows are
produced  by the same mechanism. This is a recurrent characteristic
in all the best-studied cases.  

The characteristics and the location of the outflow in 3C~305 --- but
also in other cases like 3C~293 and IC~5063 (Morganti et al. 2007) ---
strongly suggests that {\em the outflow is produced by the interaction 
between the radio jet and the ISM}.  

As final remark, it is important to note that the presence of cold gas
associated with the observed fast  outflows indicates that  the gas
can cool efficiently after a strong interaction, as predicted by
numerical simulations of jets impacting on gas clouds  
(Mellema et al. 2002; Fragile et al. 2004).  

\section{Implications}

Fast gaseous outflows of ionized and neutral atomic gas (broad
blueshifted \HI\ absorption) are extremely  common  in young, powerful
radio sources. In addition, we also note that recently
restarted radio sources (such as, e.g., 3C~236 and 3C~293) show similar
characteristics. Interestingly, all the objects with fast gaseous
outflows typically have a particularly rich ISM; 
 molecular gas (CO and  H$_2$) as well as  PAH lines
detected from {\em Spitzer} IRS spectra appear to be more common in these
young sources than in large radio galaxies (Dicken et al., in preparation).
These characteristics support the idea that these objects have
recently been triggered (or re-triggered) by a gas-rich merger/accretion
and the newly born (or re-born) radio jet is  now fighting its way
through the medium deposited by this event.  

Puzzling in this respect is the result from  the detailed study of
the line ratios of the ionized gas presented by Holt et
al. (2009). This study shows that the ionization of these lines is not
dominated by shocks, as one would have expected from their
kinematics. While densities  and temperatures derived from the broad
emission-line components (i.e., 
the components that are more kinematically disturbed by the
passage of the radio plasma) are typically high, as expected in the
presence of fast shocks, Holt et al. (2009) were  unable to
unambiguously distinguish the dominant ionization mechanism using the
optical emission-line ratios.  

If gas outflows are so ubiquitous in young, powerful radio sources,
what is the  impact they have on the evolution of the host galaxy? 
The results and the picture  sketched above are consistent with the
idea that the interaction between the radio plasma and the ISM would
actually help clear the central regions of material deposited
during the event (e.g., interaction/merger) that could have
triggered both AGN and starbursts. In this phase, the black hole  is
also rapidly growing (close to the Eddington rate), while the AGN is
obscured by the large concentration of dust and gas. The effect of the
outflows should therefore be to remove gas from the central regions and
to halt both star formation and further accretion. According to the
models, this would require that $\sim\! 5$--$10$\% of the accretion power
of the AGN drives the winds/outflows (Di Matteo et al. 2005; Hopkins et
al. 2005).  

It is, therefore, interesting to see what is the mass of the gas
outflow that we observe in the young radio galaxies studied so far.
From the best-studied case PKS 1549--79 (see Holt et al. 2006), 
we derive for the ionized gas:
\begin{itemize}
\item Mass of ionized gas in the outflow: $1.9\times10^4\,M_\odot <
M_{\rm outflow} < 1.9\times10^6\,M_\odot$.
\item Mass outflow rate: $0.12 <  \dot{M} < 12\,M_\odot\,{\rm yr}^{-1}$.
\item Energy flux: $5.1\times10^{40}\,{\rm  erg\ s}^{-1}
< \dot{E} < 5.1\times10^{42}\,{\rm  erg\ s}^{-1}$.
\item Kinetic power: $1.6\times 10^{-6} < 
\dot{E}/L_{\rm edd} < 1.6 \times 10^{-4}$.
\end{itemize}

From these numbers, it is clear that only a small
fraction of the accretion power drives the warm outflow in PKS
1549--79 (Holt et al. 2006). Thus, despite the evidence for rapid
outflows in the warm gas, the estimated kinematic power in the warm
outflow is several orders of magnitude less than required by the
feedback models. One possible explanation for this apparent
discrepancy is that much of the mass of the AGN-induced outflow is
tied up in cooler or hotter phases of the interstellar medium.

The \HI\ component of the outflows appears to be systematically more
prominent:
\begin{itemize}
\item Mass of neutral gas in the outflow: $\sim\! 10^6\,M_\odot$.
\item Mass outflow rate: up to $\sim\! 50\,M_\odot\ {\rm yr}^{-1}$.
\end{itemize}
These parameters place the neutral outflows in the same range as
``mild'' starburst-driven superwinds in ULIRGs (Rupke et al. 2002).
However, also in the case of \HI\ outflows, the ratio between the
kinetic powers associated with these gaseous outflows and the
Eddington luminosity of the galaxy is still only about a few $\times
10^{-4}$ and therefore relatively low compared to that required by
galaxy evolution feedback models.

\section{Conclusions}

The studies done so far clearly show evidence for AGN-induced outflows
on all scales in radio galaxies and in particular in young, powerful
radio sources.  We find that in young (or restarted) radio galaxies
($< 5$ kpc scale) there is almost ubiquitous evidence for nuclear warm
outflows, but they are much less powerful than required by feedback models for
bulge evolution.  Massive neutral outflows are also detected; these
are more massive, comparable to the ``mild'' starburst-driven winds
detected in ULIRGs.  These outflows may be common in powerful radio
sources, but they are difficult to detect with the sensitivity of
present-day radio telescopes.  The likely origin of (at least some of)
these outflows is interaction between the expanding jets and ISM.
Much of the power in  AGN-induced outflows may be associated with
hotter or cooler phases of the ISM.

As mentioned above, the radio part of these studies is now limited by
the sensitivity of present-day radio telescopes.  Fortunately, many
new radio facilities are planned or are in the process of becoming
available. Ultimately, the Square Kilometer Array (SKA, Lazio 2009)
will provide a major step forward by allowing a complete census of
\HI\ in radio sources (including the weak ones!), detailed studies
of the kinematics of the gas close to the nucleus, with the possibility
of looking for outflows in weaker sources (Morganti et al. 2004).
  
In the shorter term, other upgraded or new facilities e.g., EVLA,
ASKAP (Johnston  et al. 2008), MeerKat (Booth et al. 2009), and  Apertif
(focal plane array upgrade of the WSRT, Oosterloo et al. 2009) will
soon allow to improve on our knowledge of the \HI\ characteristics in
radio-loud sources. 
 
Furthermore,  the LOw FRequency Array (LOFAR, de Vos et al. 2009;  Morganti
et al. 2009) that is now coming on-line  and will   operate
between 10 and 200\,MHz,  will allow  detections of relic structures
(also in galaxies that are not currently radio loud!), thus providing
insight on the  duty cycle of radio activity.

\newpage


\end{document}